\documentclass[twocolumn]{aastex631}

\usepackage{hyperref}
\usepackage{amsmath} 
\usepackage{CJKutf8}
\usepackage{flushend}
\usepackage{threeparttable}
\usepackage{makecell}

\usepackage{soul}

\hypersetup{hypertex=true,colorlinks=true,
linkcolor=blue,anchorcolor=blue,citecolor=blue}

\begin{document}

\title{Propagation of GRB Relativistic Jets in AGN Disks and Its Implication for GRB Detection}

\author[0009-0000-8848-1803]{Hao-Hui Zhang\begin{CJK*}{UTF8}{gbsn} (张浩辉) \end{CJK*}}
\affiliation{Institute of Astrophysics, Central China Normal University, Wuhan 430079, China e-mail: yuyw@ccnu.edu.cn}
\author[0000-0002-9195-4904]{Jin-Ping Zhu\begin{CJK*}{UTF8}{gbsn} (朱锦平) \end{CJK*}} 
\affiliation{School of Physics and Astronomy, Monash University, Clayton Victoria 3800, Australia}
\affiliation{OzGrav: The ARC Centre of Excellence for Gravitational Wave Discovery, Australia}
\author[0000-0002-1067-1911]{Yun-Wei Yu\begin{CJK*}{UTF8}{gbsn} (俞云伟) \end{CJK*}} 
\affiliation{Institute of Astrophysics, Central China Normal University, Wuhan 430079, China e-mail: yuyw@ccnu.edu.cn}
\affiliation{Key Laboratory of Quark and Lepton Physics (Central China Normal University), Ministry of Education, Wuhan 430079, People’s Republic of China}

\begin{abstract}
The accretion disks of supermassive black holes (SMBHs) harboring in active galactic nuclei (AGN) are considered to be an ideal site for producing different types of gamma-ray bursts (GRBs).  The detectability of these GRB phenomena hidden in AGN disks is highly dependent on the dynamical evolution of the GRB relativistic jets. By investigating the reverse and forward shock dynamics due to the interaction between the jets and AGN disk material, we find that the relativistic jets can successfully break out from the disks only for a sufficiently high luminosity and a long enough duration. In comparison, relatively normal GRB jets are inclined to be choked in the disks, unless the GRBs occur near an SMBH with relatively low mass (e.g., $\sim 10^6M_{\odot}$). For the choked jets, unlike normal GRB prompt and afterglow emission, we can only expect to detect emission from the forward shock when the shock is very close to the edge of the disks, i.e., the shock breakout emission and subsequent cooling of the shock. 
\end{abstract}


\section{Introduction}

It has been proposed that within active galactic nuclei (AGN), the accretion disk around the supermassive black hole (SMBH) could harbor a large population of stars and compact objects, including white dwarfs, neutron stars (NSs), and stellar-mass BHs. There are typically two ways for a star to end up in AGN disk environments: in situ formation at the outer self-gravitating region of the disk via gravitational instability \citep{Kolykhalov1980,Shlosman1989,Goodman2004,Collin2008,Wang2011,Wang2012,Mapelli2012,Dittmann2020,Fan2023} and capture from nuclear star clusters \citep{Syer1991,Artymowicz1993,Fabj2020,MacLeod2020,WangYH2023}. These stars and compact objects can grow by accreting materials and undergo orbital migration
in extreme astrophysical environments of the AGN disks 
\citep[e.g.,][]{Mckernan2012,Bellovary2016,Perna2021b,Jermyn2021,Wang2021,Dittmann2021,Pan2021,Kimura2021,Tagawa2022,Kaaz2023,Chen2023,Grishin2023}. Then, the gathering of a large population of stars and compact objects in the inner part of disks can lead to frequent collapse of massive stars and mergers/collisions of binaries \citep[e.g.,][]{Cheng1999,Mckernan2012,McKernan2020,Stone2017,Bartos2017,Yang2019,Yang2022,Tagawa2020,Zhu2021neutrino2,Zhu2021Thermonuclear,Zhu2021Neutron,Grishin2021,Li2021,Li2022,Ren2022,LiGP2022,Li2023,Zhang2023,Luo2023,Prasad2023,Ryu2023,Liu2024}. In principle, these catastrophic explosive events could be finally detected as transient emission in some particular electromagnetic bands, which could be plausibly detected by the multimessenger observation of the BBH merger event GW190521 occurring in an AGN disk \citep{Abbott2020GW190521,Graham2020}.

Because of the dense gas environment, the AGN disk stars can experience an evolution very different from that of ordinary stars surrounded by normal interstellar medium \citep{Cantiello2021}. Specifically, it is expected that the AGN disk stars can spin up to potentially near-critical rotation, due to the accretion of substantial amounts of mass from the AGN disk, and undergo quasi-chemical homogeneous evolution to become compact hydrogen-free stars \citep{Cantiello2021,Jermyn2021}. The final state of these stars are similar to those of the progenitors of long gamma-ray bursts (lGRBs) \citep[e.g.,][]{Yoon2005,Woosley2006,Hu2023}, the collapse of which can drive a pair of relativistic jets propagating within the disk environment. Meanwhile, short GRB phenomena can also be expected to be produced by the frequent mergers of binary neutron star (BNS) and NSBH systems in the inner part of AGN disks \citep[e.g.,][]{Cheng1999,McKernan2020,Zhu2021Neutron,Perna2021b}. Furthermore, in the AGN disks, powerful jets could even be launched from solitary accreting BHs and remnant BHs after BBH mergers \citep[e.g.,][]{Wang2021BBH,RodriguezRamirez2023,Tagawa2023neutrino,Tagawa2023EMfromBH,Zhu2023,Chen2024}.

Simply according to the knowledge of normal GRBs, it is suggested that the relativistic jets launched in AGN disks could be able to produce GRB prompt emission \citep{Perna2021,Lazzati2022,Yuan2022,Ray2023} and afterglow emission \citep{Perna2021,Wang2022,Kathirgamaraju2023}. Recently, a long GRB event was discovered to be located at the nucleus of its galaxy host \citep{Levan2023}, which somewhat confirmed the above expectation and increased the interest of this topic \citep{Levan2023,Lazzati2023}. However, by considering of the high density of AGN disks, the dynamical evolution of relativistic jets in the disks should be very different from that of normal GRBs and, thus, different electromagnetic outcomes can be expected.  For example, \cite{Zhu2021Neutron,Zhu2021neutrino1} found that relativistic jets could be easily choked in AGN disks, in which case the primary emission signature could be the breakout emission of the shock driven by the chocked jets and accompanied high-energy neutrino emission. 

The dynamical evolution of relativistic jets in AGN disks can undergo three stages. In Stage I, the jet launched from an energy engine penetrates through the progenitor material (e.g., the stellar envelope for lGRBs and merger ejecta for sGRBs), during which the jet can be collimated \citep[e.g.,][]{Matzner2003,Bromberg2011,Yu2020}. In Stage II, the jet lastingly interacts with the disk medium, driving a forward shock moving into the medium and a reverse propagating into the jet material. As a result, a cocoon component could be formed to surround the jet if the disk medium is dense enough, just as the situation of Stage I. In Stage III, the reverse shock could cross the entire jet before the head of the jet reaches the surface of the AGN disk. In this case, we regard the jet is chocked in the disk. Afterward, the forward shock can continuously move in the disk medium and the shocked material gradually expands laterally. We can use the term ``jet remnant" to refer to the combination of the expanding choked jet and swept-up disk material, in order to distinguish it from the relativistic jet. Finally, the jet remnant can approach the outer edge of the disk and break out from it. While Stage III had been investigated carefully before \citep{Zhu2021Neutron}, the dynamical evolution in Stage II is however usually ignored in previous studies \citep[e.g.,][]{2022Wang, 2023Kathirgamaraju,2024Huang}.

Therefore, in Section \ref{sec:JetDynamics}, we describe the dynamical model of relativistic jets in AGN disks taking into account the influence of the reverse shock. Numerical calculation results are presented in Section \ref{sec:results}, including the parameter dependence of the results and the comparison with the previous works. A summary is given in Section \ref{sec:conclusion}.

\section{The Model} \label{sec:JetDynamics}
\subsection{Disk Structure}

\defcitealias{Sirko2003}{SG}\defcitealias{Thompson2005}{TQM} 

\begin{figure*}[]  
\centering
\includegraphics[width=0.325\textwidth]{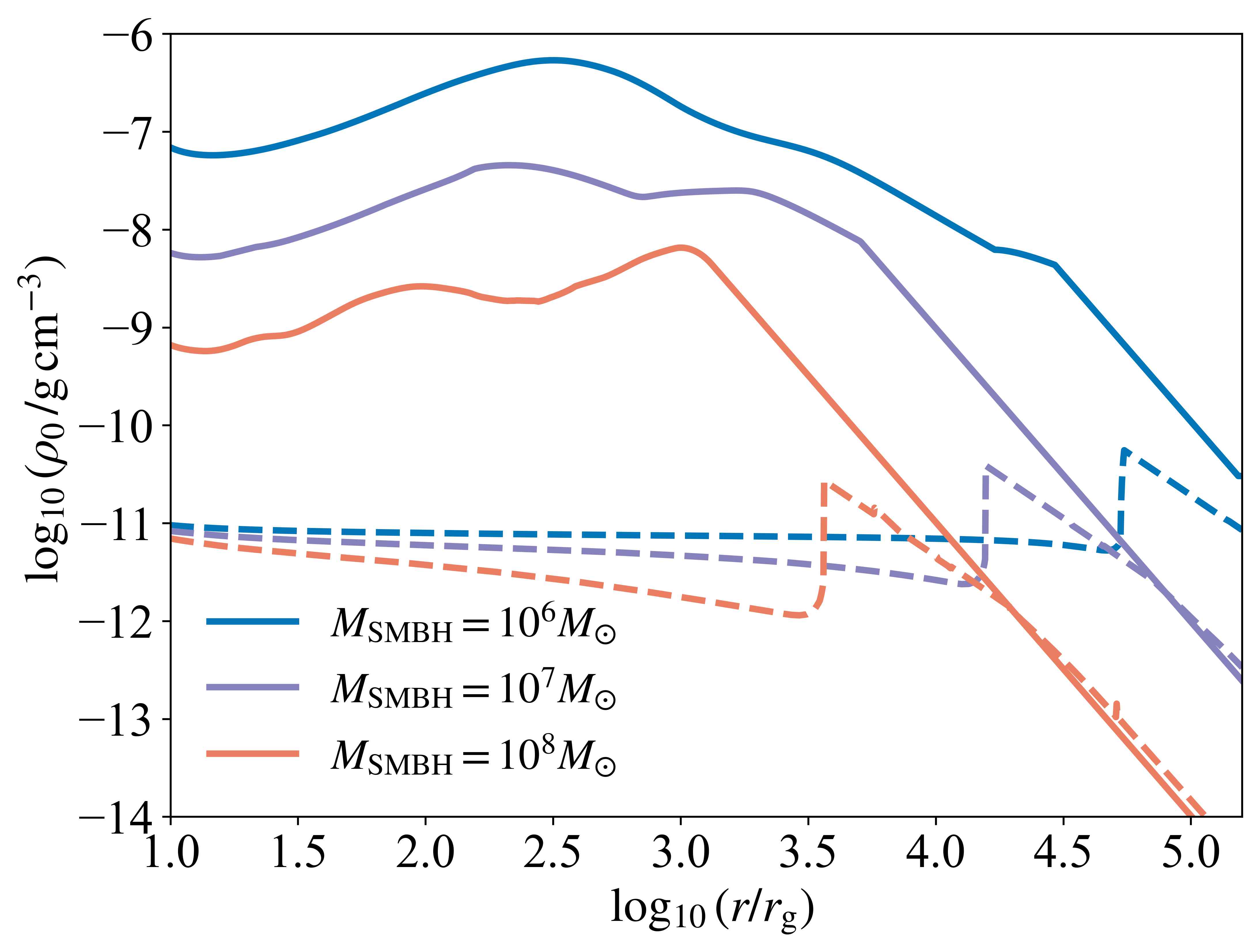} 
\includegraphics[width=0.315\textwidth]{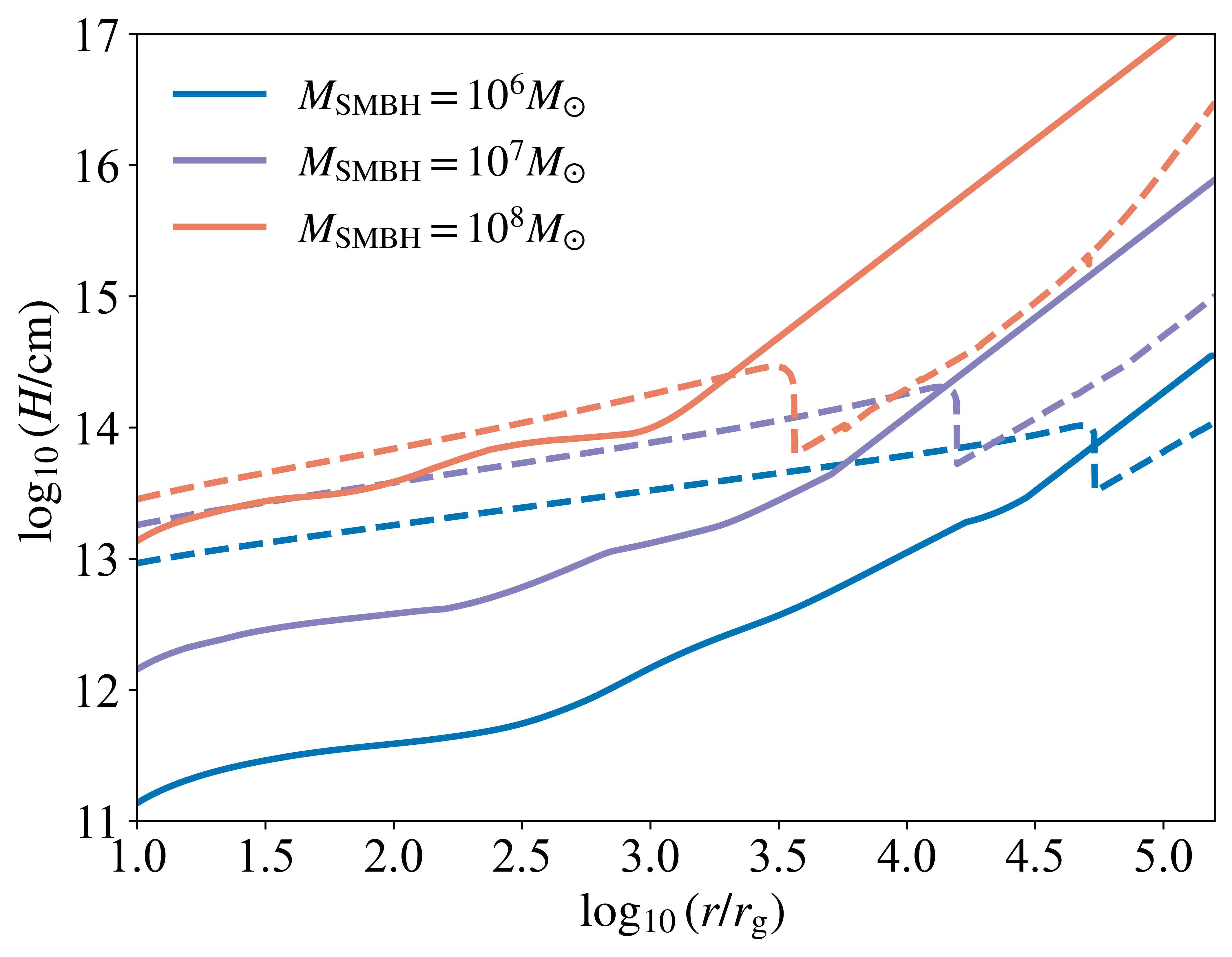} 
\includegraphics[width=0.318\textwidth]{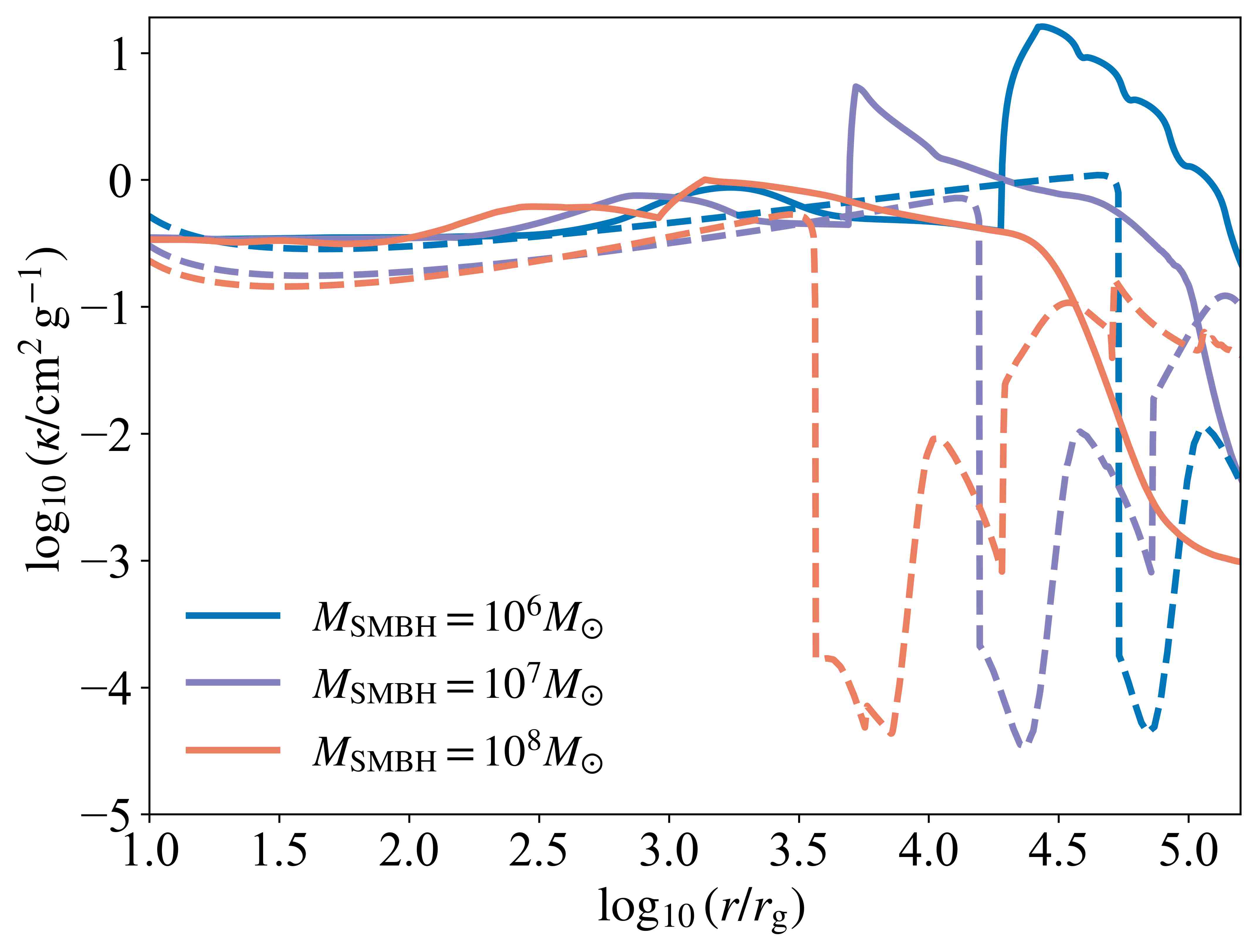} 
\caption{Mid-plane density $\rho_0$ (left panel), disk scale height $H$ (middle panel), and opacity $\kappa$ (right panel) of AGN disks as a function of the radial distance $r$ to the SMBH. The solid and dashed lines corresponds to the \citetalias{Sirko2003} and \citetalias{Thompson2005} disk models, respectively.} \label{Figure 1}
\end{figure*}

In our calculations, we consider two specific AGN disk models from \cite{Sirko2003} and \cite{Thompson2005}, abbreviated as the \citetalias{Sirko2003} model and \citetalias{Thompson2005} model, respectively. The \citetalias{Sirko2003} model supposes that angular momentum transport in the AGN disk proceeds via local viscosity, and the outer disk regions can be gravitationally stable, maintained by the heat from embedded stars, while angular momentum transport via global torques throughout the disk is assumed in the \citetalias{Thompson2005} model. In Figure \ref{Figure 1}, we show some critical physical parameters of these two disk models, including the mid-plane density $\rho_0$, disk scale height $H$, and opacity $\kappa$, as functions of the radial distance away from the central SMBH $r$ in units of the SMBH gravitational radius $r_{\rm g}$. Assume that the disk pressure is always determined by the gas pressure, the disk vertical density profile can be \citep[e.g.,][]{Netzer2013}
\begin{equation}
     \rho_{\mathrm{d}}(r,z)=\rho_{\mathrm{0}}(r)\mathrm{exp}\left[-\frac{z^{2}}{2H(r)^{2}}\right],
\end{equation}
where $z$ is the vertical height. In addition, we define the interstellar medium density (ISM) of $\rho_{\rm ISM}\sim10^{-25}\,{\rm g}\,{\rm cm^{-3}}$ as the floor density of AGN disks. The boundary between the disk and ISM $z_\star(r)$ is assumed to be the disk surface.

\subsection{Dynamics of Reverse-Forward Shocks} \label{sec:JetPropagation}
The dynamics of a jet propagating through a dense medium has been widely investigated both analytically and numerically in the literature, specifically for a stellar envelope after the supernova explosion of massive stars or an ejecta from NS mergers \citep[e.g.,][]{Matzner2003,Bromberg2011,Nagakura2014,Bromberg2016,Mooley2018,Geng2019,Hamidani2020,Yu2020}.
When a jet propagates through a stationary AGN disk atmosphere, the dynamical evolution of the jet head can be determined by the mechanical equilibrium between the shocked disk medium and shocked jet as \citep{Matzner2003,Matsumoto2018} 
\begin{equation}\label{headEq}
 \rho_{\rm j} h_{\rm j} \Gamma_{\rm j}^{2} \Gamma_{ \rm h}^{2}\left(\beta_ {\rm j}-\beta_{\rm h}\right)^{2} + P_{\rm j}=\rho_ {\rm d} h_{\rm d} \Gamma_{\rm h}^{2} \beta_{\rm h}^{2} + P_{\rm d},
\end{equation}
where $\rho$, $P$, and $h=1+P/\rho{c}^2$ are the density, pressure, and dimensionless specific enthalpy, measured in the fluid rest frame, while $\beta$ and $\Gamma$ are the dimensionless velocity and Lorentz factor, respectively, with the subscripts ``d," ``j," and ``h" representing the regions of the unshocked disk material, unshocked jet, and jet head. In the framework of GRB, the unshocked jet and unshocked disk medium are both considered to be cold, indicating that $P_{\rm j}\approx0$ and $P_{\rm d}\approx0$. The critical parameter $\widetilde{L}$, which determines the evolution and collimation of the jet, is defined as the dimensionless ratio between the jet energy density and rest-mass energy density of AGN disk material, i.e.,
\begin{equation}
\widetilde{L}\equiv\frac{\rho_{\rm j}h_{\rm j}\Gamma_{\rm j}^2}{\rho_{\rm d}}\simeq{L_{\rm j}\over\Sigma_{\rm h}\rho_{\rm d}c^3},
\end{equation}
where $L_{\rm j}$ is the jet luminosity, $\Sigma_{\rm h}$ is the cross-section of the jet head, and $c$ is the speed of light. Then, $\beta_{\rm h}$ can be derived to \citep{Matzner2003,Bromberg2011}
\begin{equation}
    \beta_{\rm h}\approx{\beta_{\rm j}\over 1+\widetilde{L}^{-1/2}},\label{2}
\end{equation}
where $\beta_{\rm j}\approx 1$. Therefore, the jet head can have a relativistic velocity if $\widetilde{L}\gg1$; otherwise, it is mildly relativistic or non-relativistic. The distance of the jet head moving in the disk is given by $z_{\rm h}=\int_0^t\beta_{\rm h}cdt$. Correspondingly, the cross section of the jet head can be given as \citep{Komissarov1997,Bromberg2011}
\begin{equation}
\Sigma_{\rm h}\approx\pi\theta_{\rm j}^{2}\times\left\{ 
\begin{array}{lc}
z_{\rm h}^{2} , & {~\rm for~}  z_{\rm h}< {\hat{z}}/{2}\\
\left({\hat{z}}/{2}\right)^{2} , & {~\rm for~}  z_{\rm h}\geq {\hat{z}}/{2}
\end{array}\right.
\end{equation}
where $\theta_{\rm j}$ is the initial opening angle of the jet, $\hat{z}\approx({L_{\rm j}}/{\pi cP'_{\rm c}})^{1/2}$ is the radius that the shock converges, and $P'_{\rm c}$ is the comoving pressure of the cocoon material with the prime representing that the quantity is measured in the comoving frame of the jet head. The opening angle could decrease with time if cocoon pressure is high enough to collimate the unshocked jet.

The mass and energy injected into the cocoon can be estimated by 
\begin{equation}
M_{\rm c}=\int_0^t\eta \Sigma_{\rm h}\rho_{\rm d}(z_{\rm h})\beta_{\rm h}cdt
\end{equation}
and
\begin{equation}
E_{\rm c}=\int^{t}_{0}\eta L_{\rm j}(1-\beta_{\rm h})dt,
\end{equation}
respectively, where $\eta=\min(2/\Gamma_{\rm h}\theta_{\rm h}, 1)$ represents the fraction of the mass and energy that can flow into the cocoon \citep{Bromberg2011}, $t$ is the dynamical time in the local rest frame, and $\theta_{\rm h}=\min\left[\theta_{\rm j},(\Sigma_{\mathrm{h}}/\pi z^{2}_{\mathrm{h}})^{1/2}\right]$ is the opening angle of the jet head relative to the progenitor center. By assuming a cylinder shape for the cocoon, one can estimate its volume to be $V_{\rm c}=\pi r_{\rm c}^2z_{\rm h}/\Gamma_{\rm h}^2$. Then, the comoving density and pressure of the cocoon can be expressed as
\begin{equation}
\bar{\rho}'_{\rm c}={M_{\rm c}\over V_{\rm c}\Gamma_{\rm h}}\approx {\Gamma_{\rm h}M_{\rm c}\over \pi r_{\rm c}^2z_{\rm h}}
\end{equation}
and
\begin{equation}
P'_{\rm c}={E_{\rm c}/\Gamma_{\rm h}\over 3V_{\rm c}\Gamma_{\rm h}}\approx {E_{\rm c}\over 3\pi r_{\rm c}^2z_{\rm h}},
\end{equation}
which yield $\beta_{\rm c}'=\sqrt{{P'_{\rm c}}/{\bar{\rho}'_{\rm c}(z_{\rm h})c^{2}}}\approx\sqrt{E_c/(3\Gamma_{\rm h}M_{\rm c}c^2)}$. As a result, one can obtain the lateral radius of the cocoon, $r_{\rm c}=\int_0^{t/\Gamma_{\rm h}}\beta'_{\rm c}cdt'$, and the opening angle of the cocoon relative to the progenitor center, $\theta_{\mathrm{c}}= r_{\mathrm{c}}/z_{\mathrm{h}}$.

The duration of the reverse shock $t_{\rm cr}$ can be defined by the following equation
\begin{equation}
    \int^{t_{\rm cr}}_{0}(1-\beta_{\rm h})dt=t_{\rm j},
\end{equation}
at which the reverse shock crosses the entire jet. Here, the duration time of the jet $t_{\rm j}$ is defined in the burst origin frame. The difference between $t_{\rm cr}$ and $t_{\rm j}$ is primarily due to the Dopper effect. If the jet head can arrive at the suface of the AGN disk before the crossing time, then the jet can successfully break out from the disk to produce normal GRB prompt and afterglow emission. Otherwise, the jet is considered to be choked.

In any case, the most concerned question is what we can observe from a propagating GRB jet. In fact, no matter there is a reverse shock or not, the forward shock can always propagate into the disk material and continuously convert its kinetic energy to heat. If this shock heat can be released successfully, then afterglow emission can be expected as usual \citep[e.g.,][]{2002Perna,Wang2022,2023Kathirgamaraju}. However, in view of the high density of the AGN disk, it needs to be noticed that the heat actually cannot escape freely from the disk before the shock reaches the height of $z_{\rm bo}$ that is determined by the following equation: 
\begin{equation}
\tau(z_{\rm bo})=\int^{z_{\star}}_{z_{\rm bo}}\kappa \rho_{\rm d}(z) dz ={c\over v(z_{\rm bo})}.\label{equation12}
\end{equation}
The above equation is written by equating the diffusion timescale of photons from $z_{\rm bo}$ to infinity to the dynamical time of the shock propagation. It is indicated that, on the one hand, only close to and beyond $z_{\rm bo}$, the heat deposited in the shocked region can start to escape from the disk, leading to the so-called shock breakout (SBO) emission. On the other hand, before the SBO, the shocks are probably radiation-mediated rather than collisionless, in which case particle acceleration by the shocks can be suppressed seriously and, thus, nonthermal emission cannot be produced.

\subsection{Dynamics after choking} \label{sec:CocoonRemnant}
For a choked jet, its emission properties after the SBO emission are dependent on the further dynamical evolution of the jet remnant consisting of the entire jet and swept-up disk material. Due to the steep decay of the disk density at the edge of the disk, it is actually not easy to precisely describe the shock dynamics during this period. Here, we take a simplified model to exhibit the approach of this dynamical evolution. First of all, it is assumed that the properties of the shocked region can always be described by the shock jump condition. The evolution of the average Lorentz factor of the jet remnant $\Gamma_{\rm jr}$ can be determined by the energy conservation law as usual as \citep{Huang1999}
\begin{equation}
 \frac{d\Gamma_{\rm jr}}{dM_{\rm jr}}=-\frac{\Gamma_{\rm jr}^{2}-1}{M_{\rm j}+2\Gamma_{\rm jr} M_{\rm jr}},  \label{JRdyn}
\end{equation}
where $M_{\rm j}=L_{\rm j}t_{\rm j}/\Gamma_{\rm j}c^{2}$ is the total mass of the jet and the mass of the swept-up disk material by the forward shock can be calculated by
\begin{equation}
\frac{dM_{\rm jr}}{dz_{\rm jr}}=2\pi (1-\cos\theta_{\rm jr})z_{\rm jr}^{2}\rho_{\rm d}(z_{\rm jr}).
\end{equation}
The distance evolution of the forward shock is given by
\begin{equation}
\frac{dz_{\rm jr}}{dt_{\rm obs}}=\frac{\beta_{\rm jr}c}{1-\beta_{\rm jr}},
\end{equation}
where $\beta_{\rm jr}$ is the dimensionless velocity of the jet remnant and $t_{\rm obs}$ is the observer time. The opening angle of the jet remnant $\theta_{\rm jr}$ can increase with time due to the lateral expansion, with the increased rate of
\begin{equation}
\label{open angle}
    \frac{d\theta_{\rm jr}}{dt_{\rm obs}}=\frac{{v_{\rm s}}(\Gamma_{\rm jr}+\sqrt{\Gamma_{\rm jr}^{2}-1})}{z_{\rm jr}},
\end{equation}
where the sound velocity of the shocked medium is \citep{Huang1999} 
\begin{equation}
    v_{\rm s}=c\sqrt{{\hat{\gamma}(\hat{\gamma} -1)(\Gamma_{\rm jr} - 1)\over 1+\hat{\gamma}(\Gamma_{\rm jr} - 1)}}
\end{equation}
with $\hat{\gamma}={(4\Gamma_{\rm jr}+1)}/{3\Gamma_{\rm jr}}$. 

In our studies, the initial values of $\Gamma_{\rm jr}$, $M_{\rm jr}$, $z_{\rm jr}$, and $\theta_{\rm jr}$ are given by $\Gamma_{\rm h}(t_{\rm cr})$, $\int_0^{t_{\rm cr}} \Sigma_{\rm h}\rho_{\rm d}(z_{\rm h})\beta_{\rm h}cdt$, $z_{\rm h}(t_{\rm cr})$, and $\theta_{\rm h}(t_{\rm cr})$, respectively. When $\theta_{\rm jr}$ increases to a relatively large size, its evolution cannot be accurately described by Equation (\ref{open angle}). Instead, the lateral expansion of the jet remnant can be significantly affected by the Taylor instability \citep{2014Duffell}. Thus, we assume a cut-off angle at $\theta_{\rm jr,max}=45^\circ$.

\section{Results} \label{sec:results}

\begin{figure*}[]  
\centering
\includegraphics[width=0.4\textwidth]{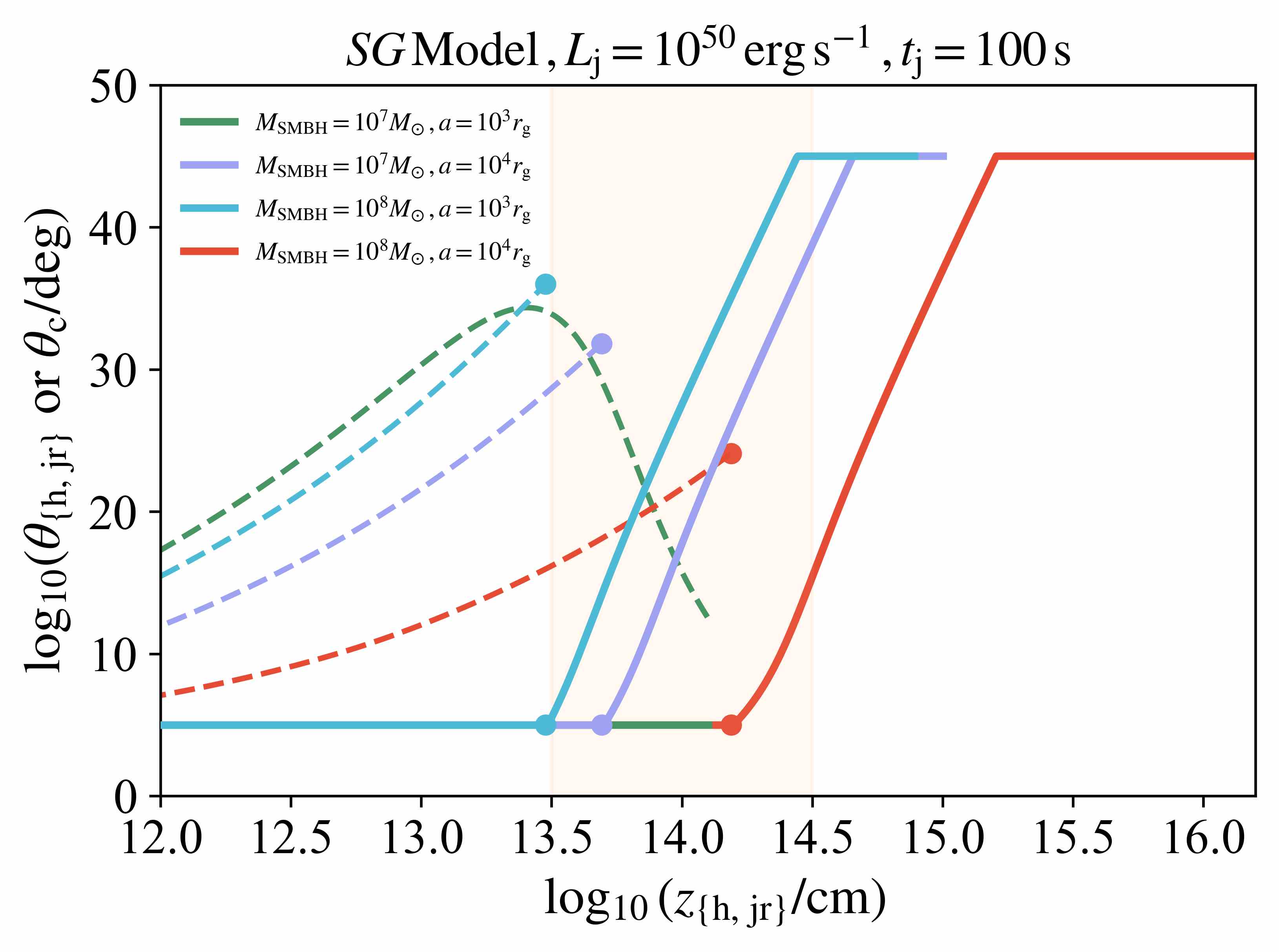} 
\includegraphics[width=0.4\textwidth]{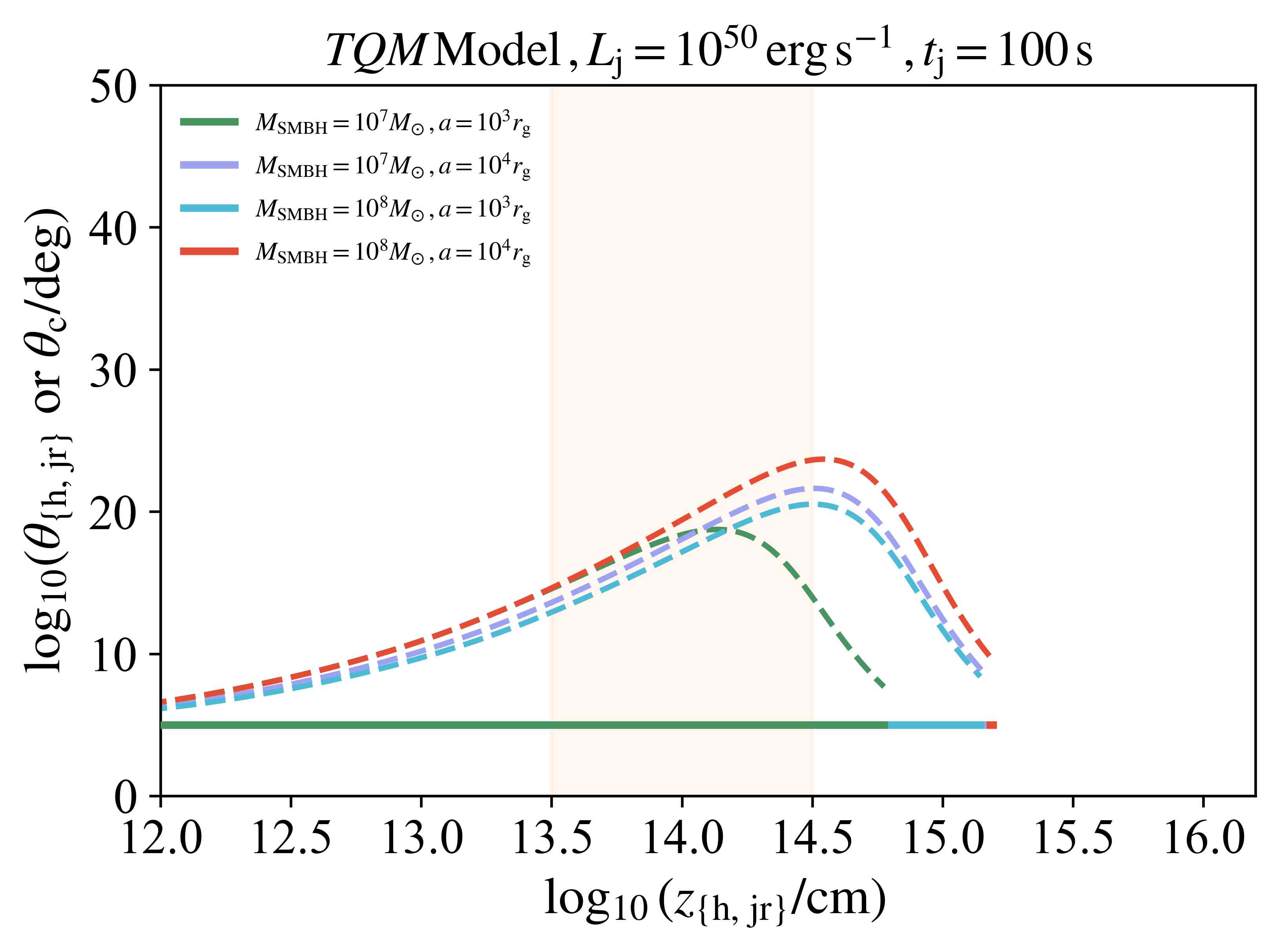} 
\caption{Evolution of the opening angles of the jet head $\theta_{\rm h}$ and jet remnant $\theta_{\rm jr}$ (solid lines), as well as the jet cocoon $\theta_{\rm c}$ (dashed lines), as functions of the motion distance. The luminosity and duration of the GRB jet are fixed as labeled. Two disk models including the \citetalias{Sirko2003} model (left panel) and \citetalias{Thompson2005} model (right panel) are considered. The locations of jet chokes are represented by solid points. }\label{angle}
\end{figure*}

\begin{figure*}[]  
\centering
\includegraphics[width=0.31\textwidth]{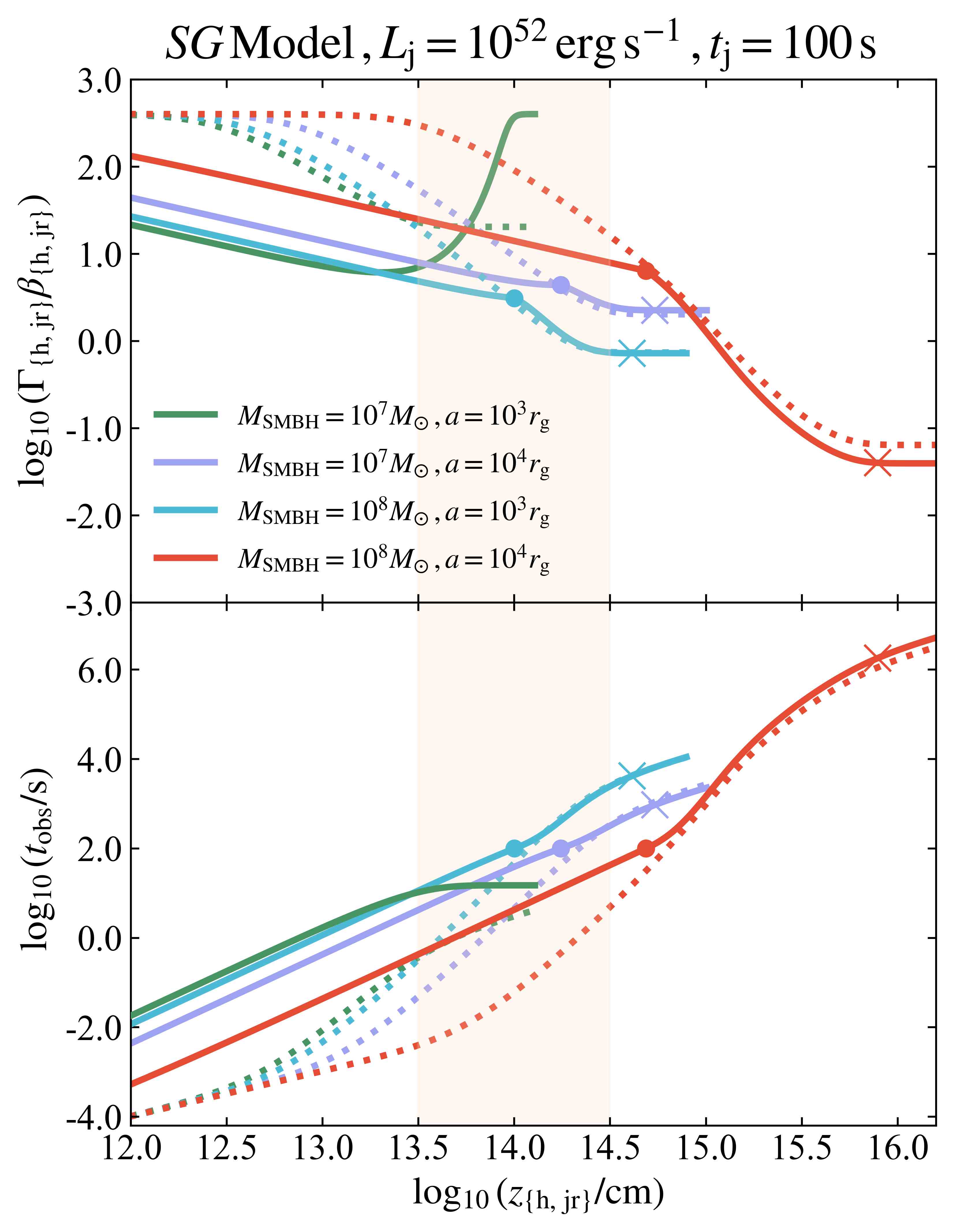} 
\includegraphics[width=0.31\textwidth]{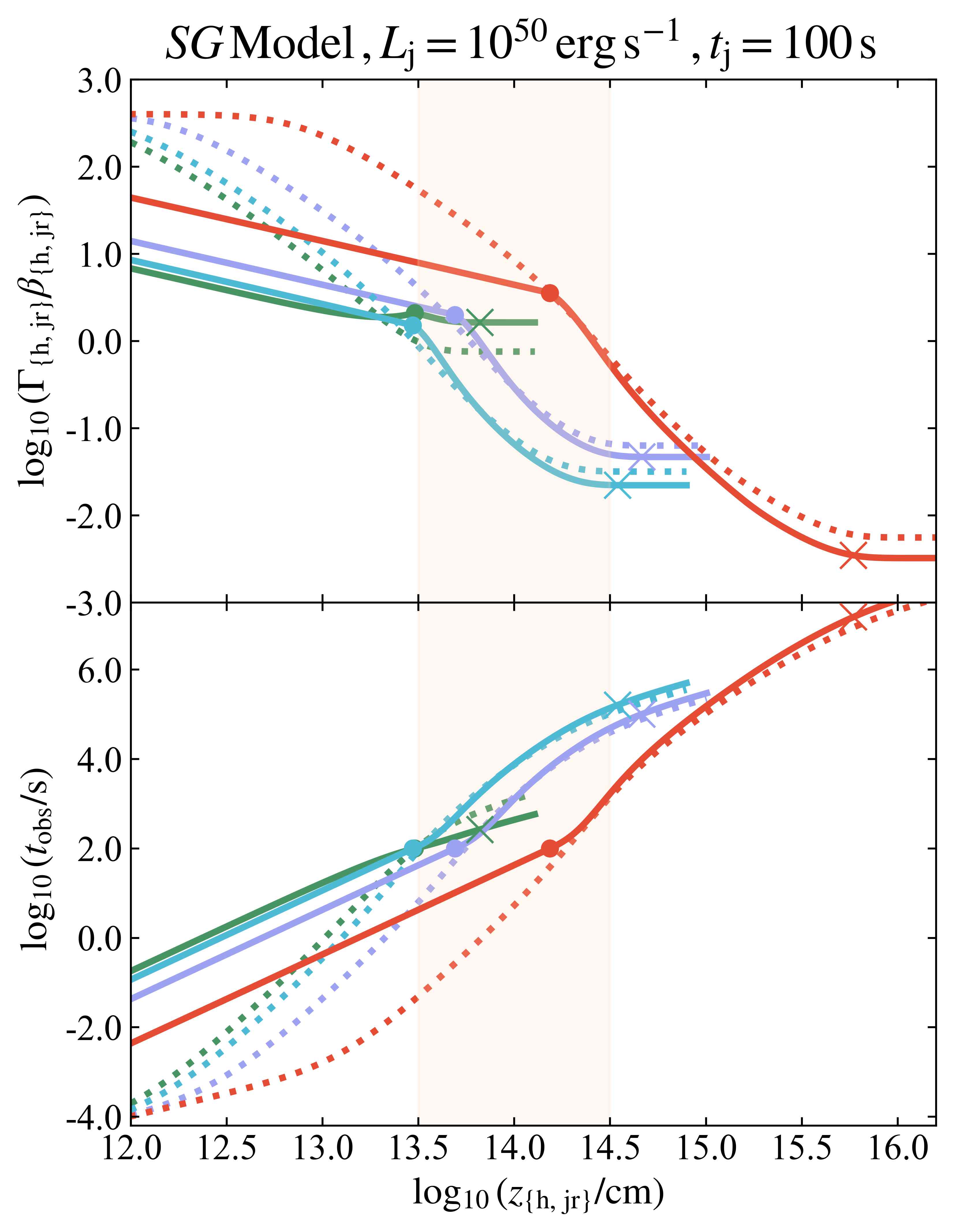}
\includegraphics[width=0.31\textwidth]{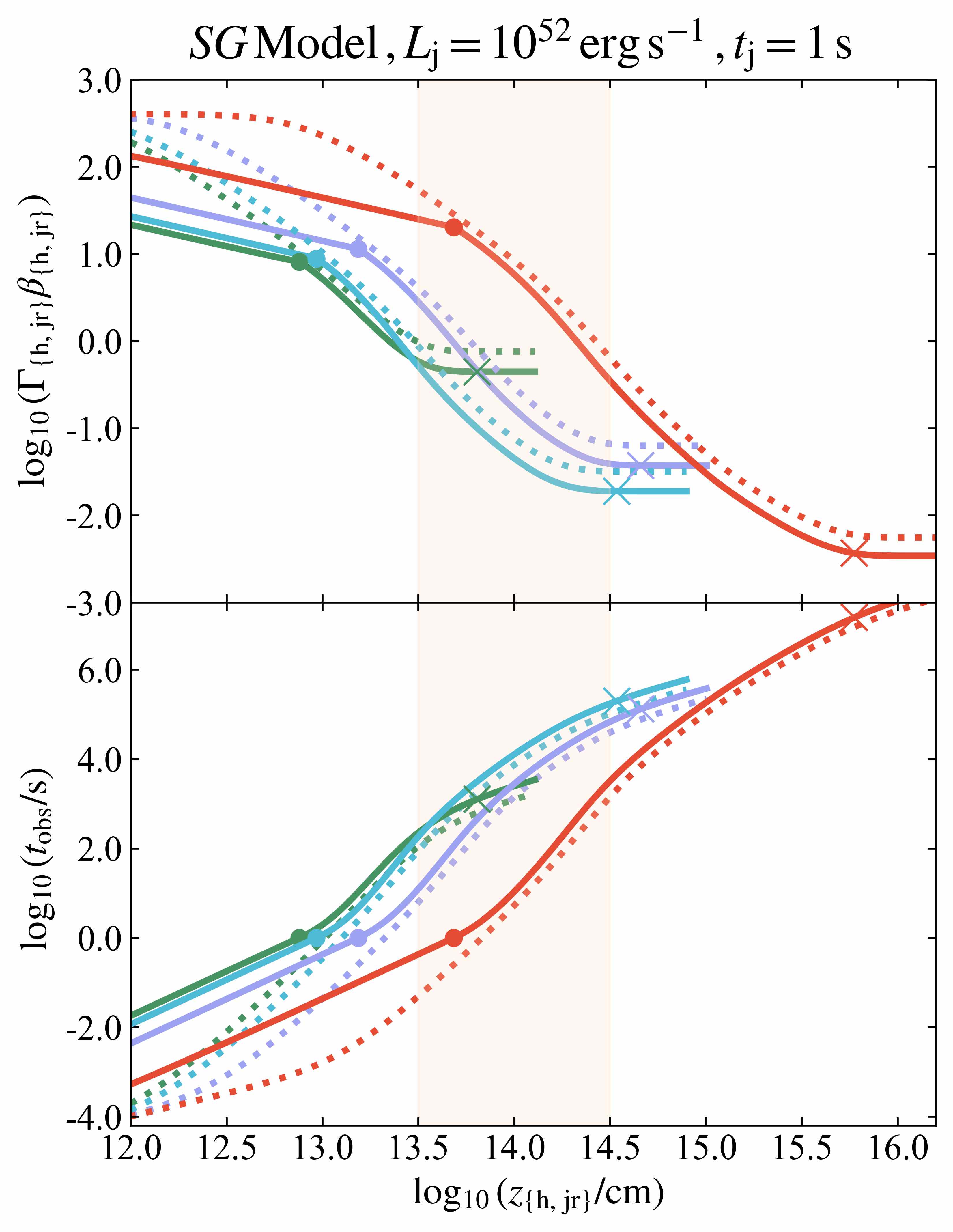}
\caption{Velocity evolution of the jet head and jet remnant with the increasing height (top panels), as well as the corresponding observer timescale of the evolution (bottom panels), for different SMBH mass and radial coordinate, considering the \citetalias{Sirko2003} disk model. The locations of jet chokes are represented by solid points, while the crosses mark the locations of shock breakouts. The observation angle $\theta_{\rm v}=0^{\circ}$ is defined. The shaded area represents the typical location of internal shocks for normal GRBs.
}\label{gauss disk}
\end{figure*}

\begin{figure*}[]  
\centering
\includegraphics[width=0.31\textwidth]{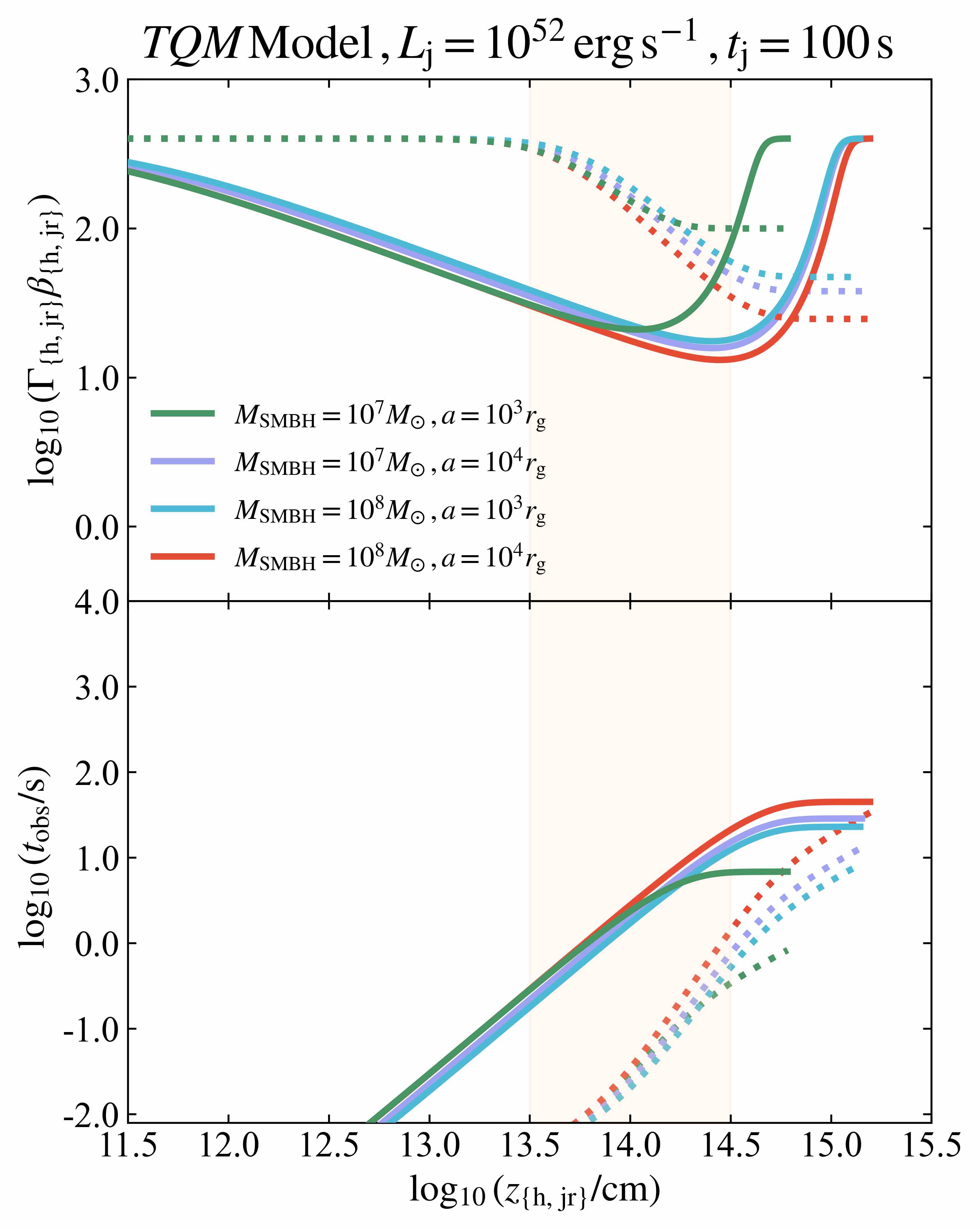} 
\includegraphics[width=0.31\textwidth]{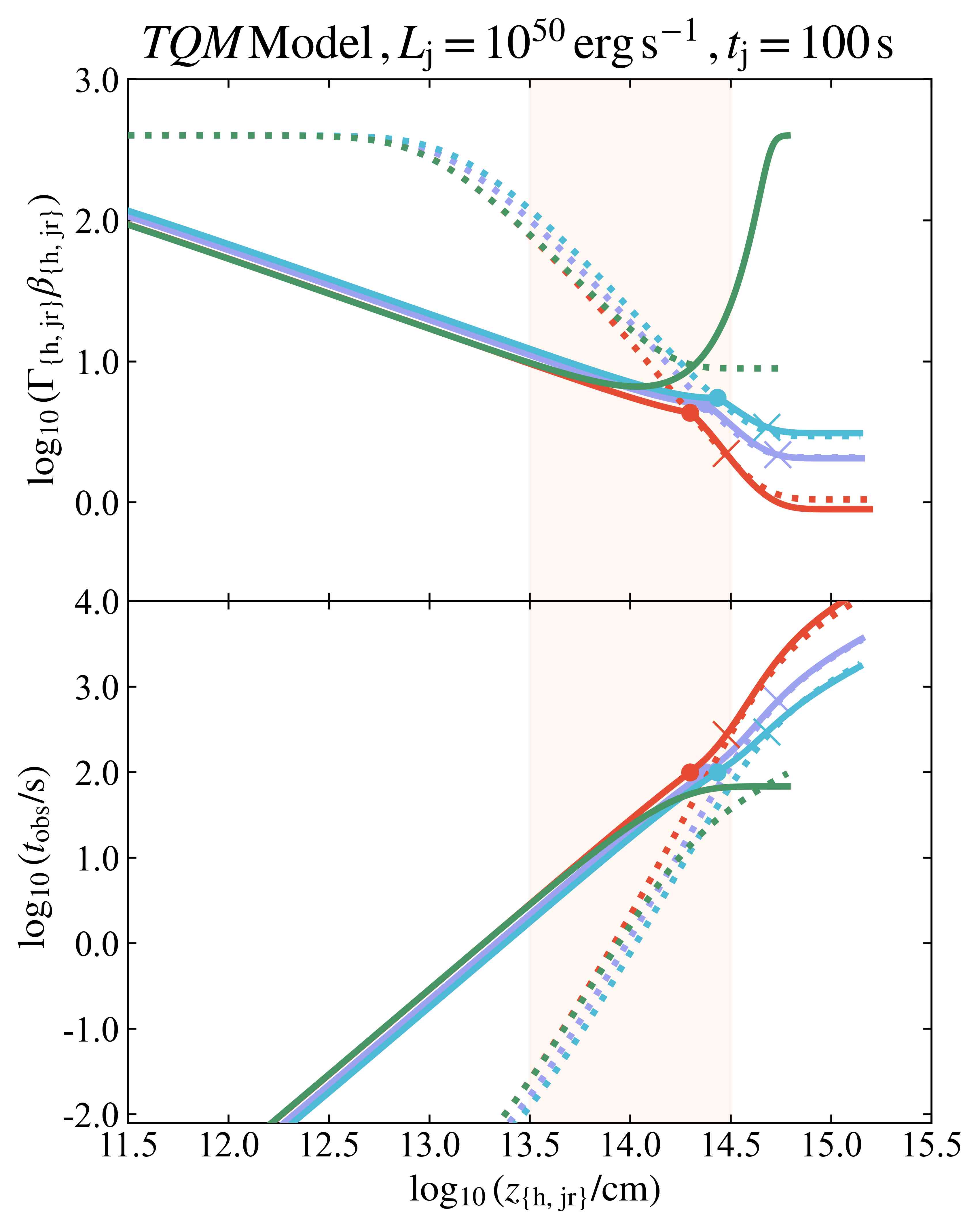} 
\includegraphics[width=0.31\textwidth]{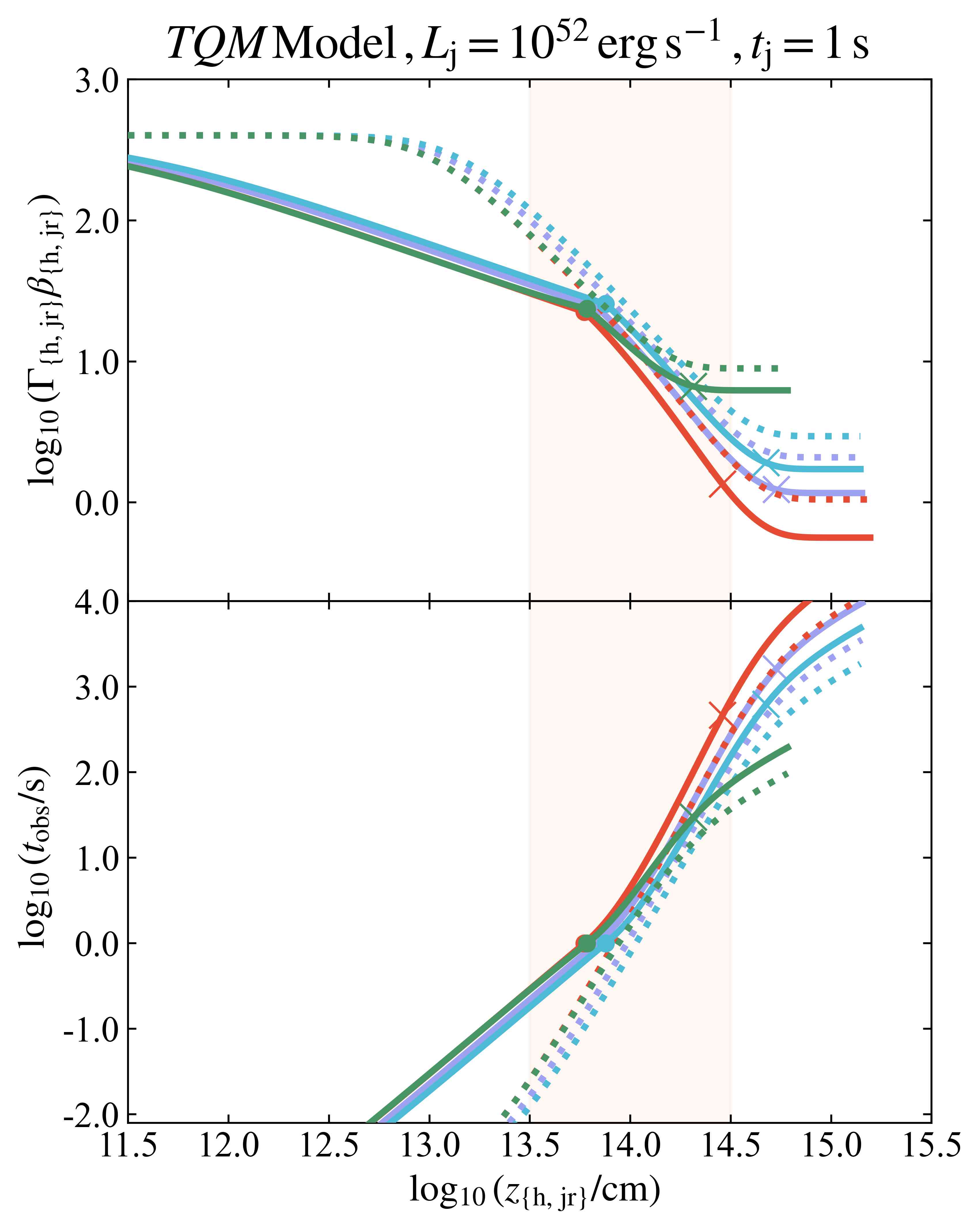} 
\caption{Similar to Figure \ref{gauss disk}, but for the \citetalias{Thompson2005} disk model.}\label{power-low disk}
\end{figure*}

Based on the above equations, we numerically calculate the dynamical evolution of the jet head and possible subsequent jet remnant. We always terminate our simulations when the jet head or jet remnant reaches to the surface of the AGN disk, defined by the floor density as $\rho_{\mathrm{d}}(r,z_*)=\rho_{\rm ISM}$. Here, for simplicity, it is assumed that GRB jets are launched in the mid-plane of AGN disks, which are always perpendicular to the disk plane. Using the burst properties of $\Gamma_{\rm j}=400$, $L_{\rm j}=10^{50}{\rm erg}\,{\rm s}^{-1}$ $t_{\rm j} = 100\,{\rm s}$, and $\theta_0=5^\circ$ and considering these bursts occur at the migration trap orbits \citep[i.e., $\sim10^3-10^5\,{r}_{\rm g};$][]{Bellovary2016,Grishin2023}, the evolution of the opening angles of the jet head/remnant and jet cocoon are shown in Figure \ref{angle}. We find that these bursts cannot be collimated by the cocoon during their propagation within the disk atmosphere, although the cocoon could become geometrically significant. Since the disk density at the migration traps is typically higher than in other regions (see Figure \ref{Figure 1}), the collimation of the jet head can always be ignored for more luminous GRB jets in the AGN disks, consistent with the results reported in \cite{Zhu2021neutrino1}. 

The velocity evolution of the jet head and jet remnant with the increasing height, as well as the corresponding observer timescale of the evolution, are presented in Figures \ref{gauss disk} (the \citetalias{Sirko2003} model) and \ref{power-low disk} (the \citetalias{Thompson2005} model) for different jet luminosities ($L_{\rm j}=10^{50}\,{\rm erg}\,{\rm s}^{-1}$ and $10^{52}\,{\rm erg}\,{\rm s}^{-1}$), durations ($t_{\rm j}=1\,{\rm s}$ and $10\,{\rm s}$), burst radial locations ($10^3\,r_{\rm g}$ and $10^4\,r_{\rm g}$), and SMBH masses ($M_{\rm SMBH}= 10^7\,M_\odot$ and $10^8\,M_\odot$). Then, on the one hand, for sufficiently high jet luminosity (i.e., $10^{52}\rm erg~s^{-1}$) and long enough duration of the central engine (i.e., $\sim100$ s), the reverse shock could persist until the jet head arrives at the disk surface. It is indicated that a tunnel could be opened in the AGN disk by the jet head. Thus, the successive relativistic jet material can finally penetrate the disk freely. In this case, both GRB prompt and afterglow emission can be detected as normal. This important result had usually been ignored in previous works since the effect of the reverse shock was not taken into account. 

On the other hand, for relatively low luminosity and short duration, the reverse shock is likely to cease when the jet head is far from reaching the disk surface. In other word, the jet is finally choked in the AGN disk. Nevertheless, even in this case, the existence of the reverse shock can still substantially influence the early dynamical evolution of the jet head, which can be clearly seen from the comparison between the solid and dotted lines. This is because the jet energy is actually transferred into the forward shock gradually through the reverse shock, rather than impulsively as considered before. Specifically, the dynamical evolution during the reverse-forward shock process can be basically determined by $\Gamma_{\rm h}^2z_{\rm h}^3\propto (1-\beta_{\rm h})L_{\rm j}t$ and $z_{\rm h}\approx ct$ (relativistic case) or $\beta_{\rm h}^2z_{\rm h}^3\propto (1-\beta_{\rm h})L_{\rm j}t$ (non-relativistic case) and $z_{\rm h}\approx \beta_{\rm h}ct$, which yields $\Gamma_{\rm h}\propto z_{\rm h}^{-1/2}$ (relativistic case) or $\beta_{\rm h}\propto z_{\rm h}^{-2/3}$ (non-relativistic case). 
\begin{figure}
     \centering
     \includegraphics[width=1\linewidth]{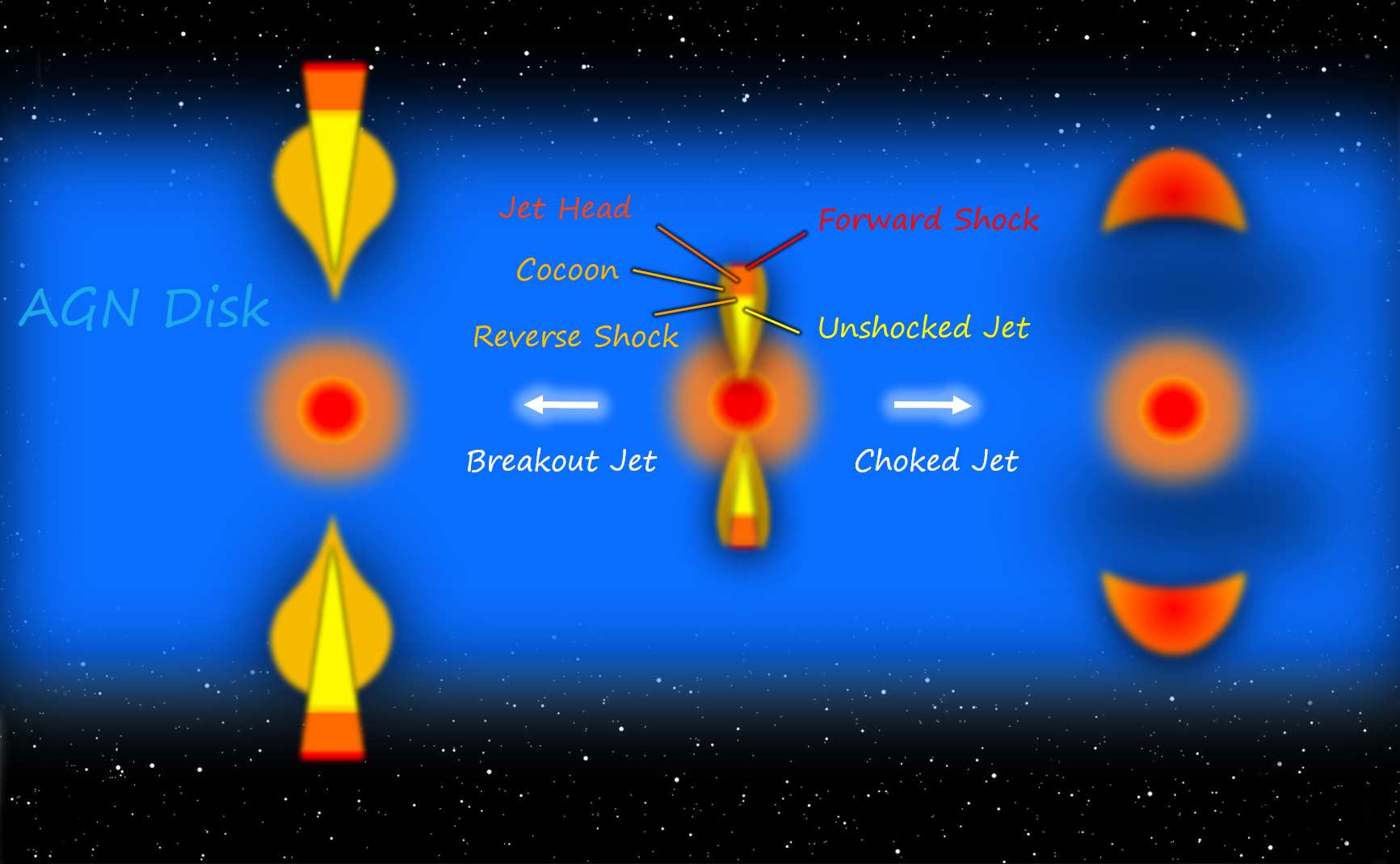}
\caption{An illustration of the different results of jet propagation in AGN disks. }\label{cartoon}
 \end{figure}
\begin{figure*}[]  
\centering
\includegraphics[width=0.45\textwidth]{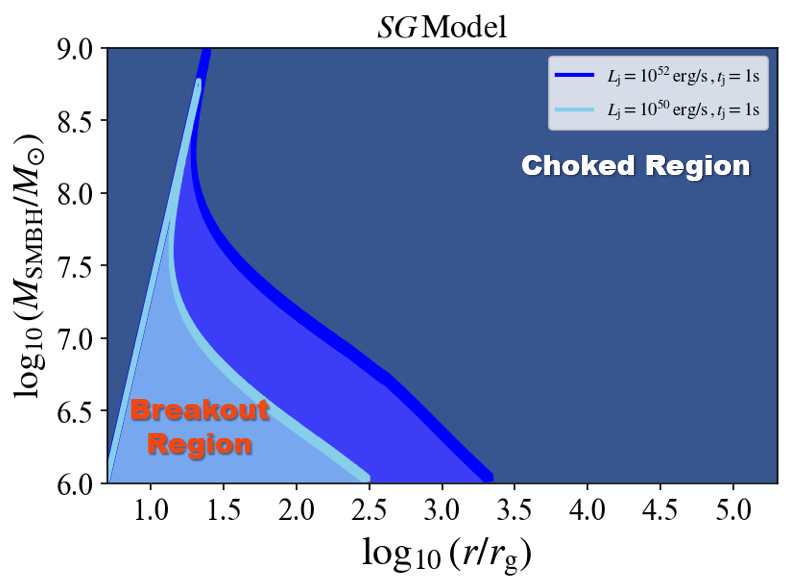} 
\includegraphics[width=0.45\textwidth]{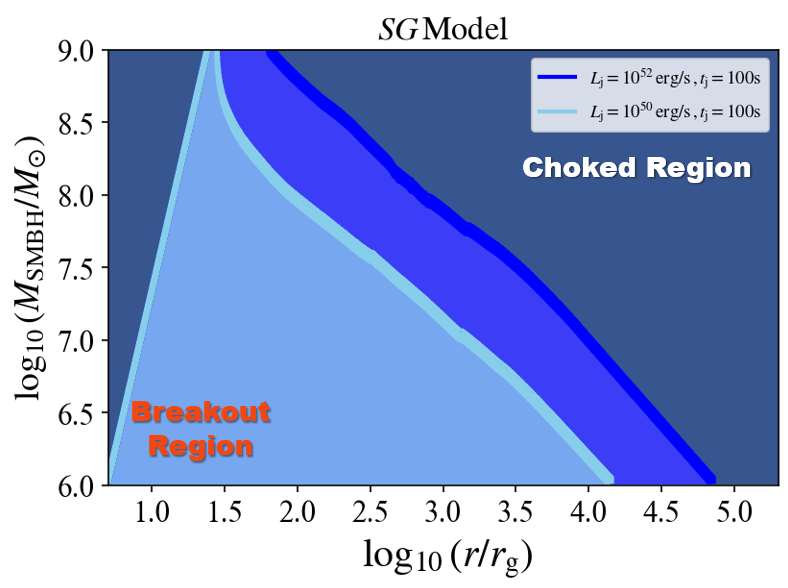} 
\includegraphics[width=0.45\textwidth]{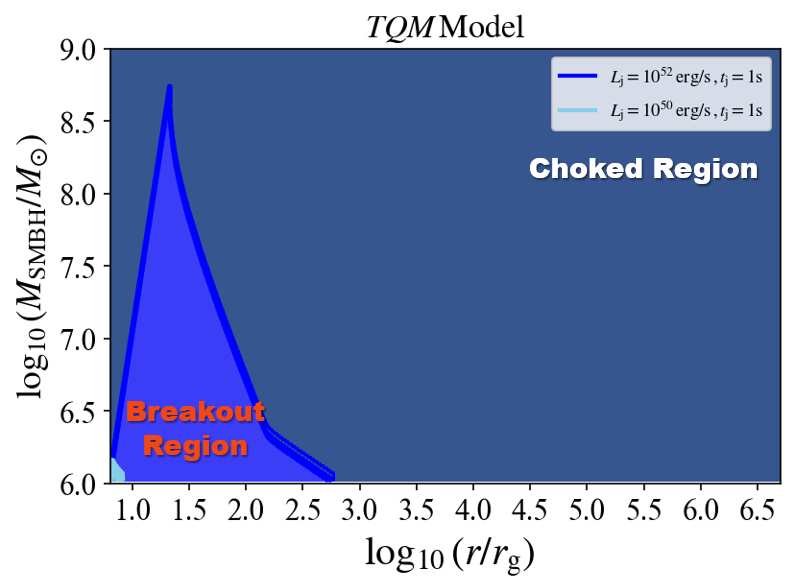}
\includegraphics[width=0.45\textwidth]{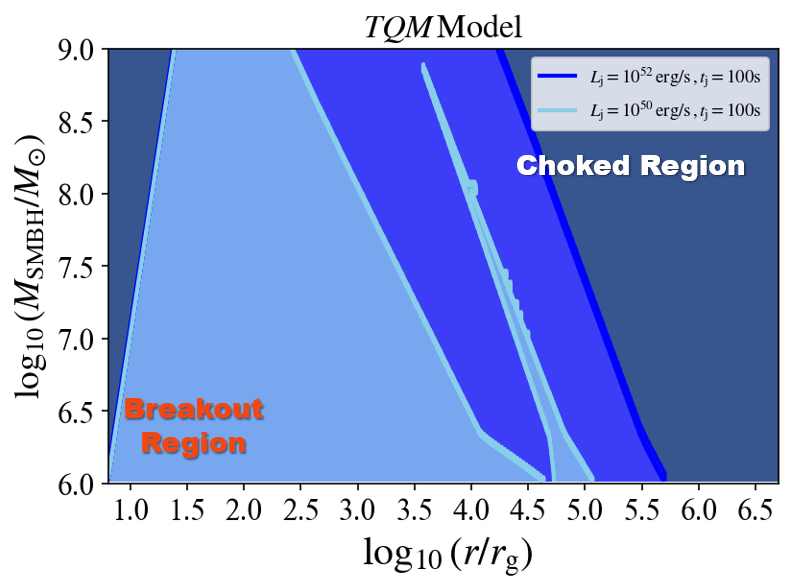}
\caption{Parameter space in the ($r, M_{\rm SMBH}$) plane for jet choking (gray region) and breakout (blue region). The top and bottom panels corresponds to the \citetalias{Sirko2003} and \citetalias{Thompson2005} disk models, respectively. The adopted jet duration and luminosity are labeled in the legend. }\label{PSboSG}
\end{figure*}

For choked jets, after the reverse shock crosses the entire jet, the dynamical evolution of the jet remnant would eventually approach to the previous results without considering the reverse shock \citep[e.g.,][]{Zhu2021Neutron}. The slight difference between the solid and dotted lines in the final stage in Figures \ref{gauss disk} and \ref{power-low disk} can arise from the facts that the jet energy is partly dissipated into the heat of the shocked jet material and, simultaneously, the lateral expansion of the shocked material can be suppressed by the cocoon at early times. In any case, the jet remnant finally becomes mild-relativistic of $\Gamma_{\rm jr}\beta_{\rm jr}\sim$ a few and even deeply non-relativistic of $\Gamma_{\rm jr}\beta_{\rm jr}\sim0.01$. According to these results, 
we can fix the position of the SBO emission as labeled by the crosses in Figures \ref{gauss disk} and \ref{power-low disk}. It can be found that, for choked jets, both the reverse shock process and the internal dissipation of the jets occurring at a typical height of $z_{\rm id}\sim10^{13.5-14.5}$ cm are actually undetectable. So, in future works, the main task is to elaborately describe the SBO emission \citep[e.g.,][]{Zhu2021Neutron} and the subsequent shock cooling and shock interaction. 

In Figure \ref{cartoon}, we show a schematic cartoon summarizing the two different dynamical outcomes of the jet, including successful breakout and choking. Strictly speaking, whether a relativistic jet can breakout from AGN disks not only depends on its luminosity and duration, but also sensitive to the distance of the GRB position to the SMBHs and the sturcture of the AGN disks. Such a dependence of jet dynamics on the parameters $r$ and $M_{\rm SMBH}$ is exhibited in Figure \ref{PSboSG}. It is showed that a relatively small mass (e.g., $\sim10^6M_{\odot}$) of the SMBHs would be beneficial for the breakout of the jets, making it possible to happen in a wide range of distance to the SMBHs, especially for the long GRBs. 

\section{Summary and Discussion} \label{sec:conclusion}

The wide existence of massive stars and compact binaries in AGN disks makes it potentially ubiquitous there to launch relativistic GRB jets. However, the detectability of these GRB phenomena is still highly dependent on whether these jets can successfully break out from the AGN disks. Therefore, in this work, we investigate the dynamical evolution of the jets by considering of the process of the reverse-forward shock interaction between the jets and the AGN disk material. It is found that only the jets with sufficiently high luminosity and sufficiently long duration could break out from the disks, leading to detectable GRB prompt and afterglow emission. Otherwise, the jets with relatively normal luminosity and duration are generally inclined to be choked in the disks. In this case, it needs to be very careful to predict their possible detection signatures.

For choked GRB jets, the primary detectable signal would be the SBO emission and, furthermore, some persistent emission could also be detected after the SBO due to the cooling of the shocked region and the continuous interaction between the jet remnant and the AGN disk material. Nevertheless, in order to describe the property of these persistent emission components, a model more elaborate that Equation (\ref{JRdyn}) is needed to describe the dynamical evolution of the forward shock, especially, when the shock propagates into the edge of AGN disks where the density declines very quickly. In contrast to the difficulty of the electromagnetic detection, future neutrino and gravitational wave detection may in principle provide alternative ways to study the GRB phenomena happening in AGN disks. For example, following \cite{Yu2020}, it can be expected that the propagation of GRB jets in the AGN disks could generate a gravitational wave memory signal, which have a characteristic frequency of $10^{-3}-10^{2}$\,Hz corresponding to the typical duration of the jets.

\begin{acknowledgements}
This work is supported by the National Key R\&D Program of China (2021YFA0718500), the National SKA Program of China (2020SKA0120300), the China Manned Spaced Project (CMS-CSST-2021-A12), the National Natural Science Foundation of China (grant No. 12393811).
\end{acknowledgements}

\bibliography{sample631}{}
\bibliographystyle{aasjournal}

\end{document}